\documentclass[prl,aps,showpacs,nofootinbib,superscriptaddress,tightenlines,twocolumn]{revtex4}

\usepackage{epsfig}
\usepackage[latin1]{inputenc}
\usepackage{float,amsmath}
\usepackage{graphicx}

\newcommand{\beq}{\begin{equation}}
\newcommand{\eeq}{\end{equation}}
\newcommand{\bqa}{\begin{eqnarray}}
\newcommand{\eqa}{\end{eqnarray}}

\def\lsim{\mathrel{\rlap{\lower4pt\hbox{$\sim$}}
    \raise1pt\hbox{$<$}}}                
\def\gsim{\mathrel{\rlap{\lower4pt\hbox{$\sim$}}
    \raise1pt\hbox{$>$}}}                

\begin{document}

\title{Photon production from an anisotropic quark-gluon plasma}

\author{Bj\"orn Schenke}
\affiliation{Institut f\"ur Theoretische Physik \\
  Johann Wolfgang Goethe - Universit\"at Frankfurt \\
  Max-von-Laue-Stra\ss{}e~1,
  D-60438 Frankfurt am Main, Germany\vspace*{5mm}}
\author{Michael Strickland}
\affiliation{Frankfurt Institute for Advanced Studies \\
  Johann Wolfgang Goethe - Universit\"at Frankfurt \\
  Max-von-Laue-Stra\ss{}e~1,
  D-60438 Frankfurt am Main, Germany \\}

\begin{abstract}

We calculate photon production from a quark-gluon plasma which is 
anisotropic in momentum space including the Compton scattering, $q(\bar{q}) \, g 
\rightarrow q(\bar{q}) \, \gamma$, and annihilation, $q \, \bar{q} \rightarrow g 
\, \gamma$, processes.  We show that for a quark-gluon plasma which has an 
oblate momentum-space anisotropy, $\langle p_L^2 \rangle < \langle 
p_T^2\rangle$, the photon production rate has an angular 
dependence which is peaked transverse to the beam line. We propose to use the 
angular dependence of high-energy medium photon production to experimentally determine 
the degree of momentum-space isotropy of a quark-gluon plasma produced in 
relativistic heavy-ion collisions.

\end{abstract}
\maketitle
\newpage


The ongoing heavy-ion experiments at the Relativistic Heavy Ion Collider (RHIC) 
and planned at the Large Hadron Collider (LHC) aim to determine information 
about QCD under extreme conditions.  The goal of these experiments is to produce 
and study the properties of a quark-gluon plasma which is expected to be 
formed when the temperature of nuclear matter is raised above its critical 
value, $T_c \sim 200$ MeV.  In the final state these experiments will generate 
on the order of $10^4$ hadronic particles.  However, during the early stages of 
the system's evolution the produced ``fireball'' has an energy density on the 
order of $20-40 \; {\rm GeV}/{\rm fm}^3$ and one should describe the system in 
terms of partonic (quasiparticle) degrees of freedom, quarks and gluons.  The 
question of whether or not the partonic degrees of freedom interact frequently 
enough to be described as an isotropic thermalized state before hadronizing and 
finally flying apart as non-interacting hadrons is an important and difficult question 
\cite{Baier:2000sb} with recent developments focusing on the impact of momentum-space
anisotropies on plasma evolution 
\cite{Mrowczynski:2000ed,Randrup:2003cw,Romatschke:2003ms,Arnold:2003rq,%
Romatschke:2004jh,Mrowczynski:2004kv,Arnold:2005vb,%
Rebhan:2005re,Romatschke:2005pm,Schenke:2006xu,Manuel:2006hg,Romatschke:2006nk,Romatschke:2006wg,%
Dumitru:2006pz}.

Absent a theoretical proof that the system will become isotropic in momentum 
space (and possibly thermalized) quickly enough, one is forced to look for 
observables which would be sensitive to early time momentum-space anisotropies 
in the quark and gluon distribution functions.  One such observable is high-energy 
medium photon production.  Because photons interact weakly with the 
surrounding matter during the evolution of the plasma and subsequent hadronic 
phase they are ideally suited for determining information about the early stages
of a heavy-ion collision.  For this reason thermal photon production from an 
isotropic quark-gluon plasma has been studied extensively 
\cite{Shuryak:1978ij,Kajantie:1981wg,Kapusta:1991qp,Redlich:1992fr,Aurenche:1998nw,%
Aurenche:1999tq,Arnold:2001ba,Arnold:2001ms}.  Thermal photons make significant
contributions to the total photon rate in the momentum range $2\;{\rm GeV}\!\lsim 
p_T\!\lsim\!6\;{\rm GeV}$. Here we are also interested in the medium produced 
photons in this momentum range; however, we relax the assumption of isotropy in order 
to see what effect this has on the angular dependence of medium photon production.

For our final results we present the energy and angular dependence of the photon 
production for two different system anisotropies.  The results show that for 
finite anisotropies photon production is peaked transverse to the anisotropy 
direction.  For oblate geometries, $\langle p_L^2 \rangle < \langle 
p_T^2\rangle$, this corresponds to production being peaked transverse to the 
beamline.  This result suggests that it may be possible to determine the time-dependent 
anisotropy of the plasma by measuring the rapidity dependence of 
photon production at various energies.


\section{Photon Rate -- Hard Contribution}
\label{sec:hardpart}

There are two hard contributions coming from the annihilation and Compton scattering diagrams.
The rate of photon production from the medium quark
annihilation diagrams can be expressed as 
\bqa
E \frac{d R_{\rm ann}}{d^3q} &=&
    {64 \pi^3 e_q^2 \alpha_s \alpha} \nonumber \\
    && \hspace{-1.2cm} \times 
        \int_{{\bf k}_1} \frac{f_q({\bf k}_1)}{k_1}
        \int_{{\bf k}_2} \frac{f_q({\bf k}_2)}{k_2}
        \int_{{\bf k}_3} \frac{1 + f_g({\bf k}_3)}{k_3} \nonumber \\
    && \hspace{-4mm} \times \, 
        \delta^{(4)}(K_1 + K_2 - Q - K_3) \left[ \frac{u}{t} + \frac{t}{u} \right]
        \; ,
\label{eq:annihilation1}
\eqa
where we have assumed massless quarks, the Mandelstam variables are defined by 
$t \equiv (K_1 - Q)^2$ and $u \equiv (K_2 - Q)^2$, $e_q^2 = 5/9$ comes from 
SU(3) Casimirs, and $\int_{\bf k} \equiv \int d^3 k/(2 \pi)^3$.  To make the 
distinction between four-vectors and three-vectors clear we use upper-case 
letters for four-vectors.  Lower-cased letters will represent the length of a 
three-vector while a bold-faced lower-cased letter stands for a three-vector. 
Here $f_{q,g}$ are the medium quark and gluon distribution functions and we 
have assumed that the distribution functions for quarks and anti-quarks are the 
same, $f_q = f_{{\bar q}}$.

In this work we will also assume that the anisotropic distribution functions are 
obtained by squeezing or stretching an arbitrary isotropic distribution along one 
direction in momentum space, i.e.,
$
f_i({\bf k},\xi,p_{\rm hard}) = 
   f_{i,\rm iso}\left(\sqrt{{\bf k}^2+\xi({\bf k}\cdot{\bf \hat n})^2},p_{\rm hard}\right) ,
$
where $i=\{ q ,g \}$, $p_{\rm hard}$ is a hard momentum scale which appears in
the distribution functions, ${\bf \hat n}$ is the direction of the anisotropy, and 
$\xi>-1$ is a parameter reflecting the strength and type of the 
anisotropy~\cite{Romatschke:2003ms}. In the following we will suppress the explicit
dependence on $\xi$ and $p_{\rm hard}$.

Anticipating the infrared divergence associated with this graph we begin by 
first changing variables in the first integration to $P \equiv K_1 - Q$ and 
introduce an infrared cutoff $p^*$ on the integration over the exchanged three-momentum 
$p$ \cite{Braaten:1991dd}. We then choose spherical coordinates with 
the anisotropy vector $\hat{\bf n}$ defining the $z$-axis.  Exploiting the 
azimuthal symmetry about the $z$-axis we also choose ${\bf q}$ to lie in the 
$x\!-\!z$ plane. Using the delta function to perform four out of the nine 
integrations and expanding out the phase-space integrals explicitly gives

\bqa
E \frac{d R_{\rm ann}}{d^3q} &=&
    \frac{e_q^2 \alpha_s \alpha}{2 \pi^6}
      \sum_{i=1}^{2}
      \int_{p^*}^\infty dp \, p^2 \,
      \int_{-1}^{1} d\cos\theta_p \, 
      \int_{0}^{2\pi} \! d\phi_p \nonumber \\
      && \hspace{-4mm} \times \;
      \frac{ f_q({\bf p}+{\bf q}) }{ \left|{\bf p}+{\bf q}\right| }
      \int_0^\infty \, dk \, k \,
      \int_{-1}^{1} \, d\cos\theta_k \; \nonumber \\
      && \hspace{-4mm} \times \;      
        f_q({\bf k}) \left[ 1 + f_g({\bf p}+{\bf k}) \right]
        \chi^{-1/2} \, \Theta(\chi) \left[ \frac{u}{t} \right] \Biggr|_{\phi_k = \phi_i} \! ,
        \nonumber \\
\label{hardannihilationrate}
\eqa
with $t = \omega^2 - p^2$, $u = (k-q)^2 - ({\bf k}-{\bf q})^2$, and $\omega = \left|{\bf p}+{\bf q}\right| - q$.
The azimuthal angles $\phi_i$ are defined through
\beq
\cos(\phi_i - \phi_p) =
    \frac{\omega^2 - p^2 + 2 k (\omega - p \cos\theta_p \cos\theta_k)}{2 p k \sin\theta_p \sin\theta_k}  \; ,
\label{eq:phiksol}
\eeq
and $\chi>0$ is given by
\bqa
\chi &\equiv& 4 p^2 k^2 \sin^2\theta_k \sin^2\theta_p - \nonumber \\
                  && \left[\omega^2 - p^2 + 2k(\omega - p \cos\theta_p \cos\theta_k)\right]^2 \, .
\eqa

The rate of photon production from the Compton scattering diagrams
can be expressed as
\bqa
E \frac{d R_{\rm com}}{d^3q} &=&
    -{128 \pi^3 e_q^2 \alpha_s \alpha} \nonumber \\
    && \hspace{-1.2cm} \times
        \int_{{\bf k}_1} \frac{f_q({\bf k}_1)}{k_1}
        \int_{{\bf k}_2} \frac{f_g({\bf k}_2)}{k_2}
        \int_{{\bf k}_3} \frac{1 - f_q({\bf k}_3)}{k_3} \nonumber \\
    && \hspace{-4mm} \times
        \delta^{(4)}(K_1 + K_2 - Q - K_3) \left[ \frac{s}{t} + \frac{t}{s} \right]
        \; ,
\label{eq:compton1}
\eqa
where we have again assumed massless quarks, $f_{\bar q}=f_q$, and the Mandelstam variables are 
defined by $s \equiv (K_1 + K_2)^2$ and $t \equiv (K_1 - Q)^2$.  Performing a 
change of variables to $P \equiv K_1-Q$ and 
proceeding as with the annihilation contribution gives

\bqa
E \frac{d R_{\rm com}}{d^3q} &=&
    -\frac{e_q^2 \alpha_s \alpha}{2\pi^6}
      \sum_{i=1}^{2}
      \int_{p^*}^\infty dp\,p^2
      \int_{-1}^{1} d\cos\theta_p
      \int_{0}^{2\pi} \! d\phi_p \nonumber \\
      && \hspace{-4mm} \times \;
      \frac{ f_q({\bf p}+{\bf q}) }{ \left|{\bf p}+{\bf q}\right| }
      \int_0^\infty \, dk \, k \,
      \int_{-1}^{1} \, d\cos\theta_k \; f_g({\bf k}) \nonumber \\
      && \hspace{-4mm} \times \;      
        \left[ 1 - f_q({\bf p}+{\bf k}) \right]
        \chi^{-1/2} \, \Theta(\chi) \left[ \frac{s}{t} + \frac{t}{s} \right] \Biggr|_{\phi_k = \phi_i} ,
        \nonumber \\
\label{hardcomptonrate}
\eqa
where here $t = \omega^2 - p^2$, $s = (\omega+q+k)^2 - ({\bf p}+{\bf q}+{\bf k})^2$, 
and recalling $\omega = \left|{\bf p}+{\bf q}\right| - q$.

The total photon production from processes which have a hard momentum exchange 
is given by the sum of Eqs.~(\ref{hardannihilationrate}) and (\ref{hardcomptonrate}).
\bqa
E \frac{d R_{\rm hard}}{d^3q} = E \left( \frac{d R_{\rm ann}}{d^3q} + \frac{d R_{\rm com}}{d^3q} \right) \, .
\label{hardphotonrate}
\eqa
To evaluate the five-dimensional integrals in Eqs.~(\ref{hardannihilationrate}) and 
(\ref{hardcomptonrate}) we use Monte-Carlo integration.
The total hard contribution (\ref{hardphotonrate}) has a logarithmic infrared 
divergence as $p^* \rightarrow 0$. This logarithmic infrared divergence will be 
cancelled by a corresponding ultraviolet divergence in the soft contribution.


\section{Photon Rate -- Soft Contribution}
\label{sec:softpart}

We now turn to the calculation of the previously excluded contribution involving soft 
momentum exchange. We adopt the Keldysh formulation of quantum field theory, 
which is appropriate for systems away from equilibrium 
\cite{Ke64,Ke65}. In this formalism both propagators and self 
energies have a $2\times 2$ matrix structure. The components $(12)$ and $(21)$ 
of the self-energy matrices are related to the emission and absorption 
probability of the particle species under consideration 
\cite{Chou:1984es,Mrowczynski:1992hq,Calzetta:1986cq}. Due to the almost 
complete lack of photon absorbing back reactions, the rate of photon emission 
can be expressed as \cite{Baier:1997xc}
\begin{equation}
 E\frac{d R_{\rm soft}}{d^3 q} = \frac{i}{2(2\pi)^3} {\Pi_{12}}_\mu^\mu (Q) \, ,
 \label{softphotonrate}
\end{equation}
from the trace of the (12)-element $\Pi_{12}$ of the
photon-polarization tensor.

We evaluate ${\Pi_{12}}_\mu^\mu$ using the hard-loop (HL) resummed 
fermion propagator derived in Ref.~\cite{Schenke:2006fz}. In that Reference we showed
that the needed off-diagonal components of the fermion self-energy 
can be expressed in terms of the retarded self-energy
\bqa 
\label{q-self} 
\Sigma(P) = \frac{C_F}{4} g^2 \int_{\bf k} \frac{f ({\bf k})}{|{\bf k}|}
            \frac{K \cdot \gamma}{K\cdot P} \,,\label{retself} 
\eqa 
where
\beq
f ({\bf k}) \equiv 2 \left( f_q({\bf k}) + f_{\bar{q}} ({\bf k}) \right) 
                   + 4 f_g({\bf k}) \; .
\eeq
Taking the HL limit where appropriate we find that
\begin{align}\label{iPi3}
 i \Pi_{12}&_\mu^\mu(Q) = -e^2e_q^2 N_c \frac{8 f_q({\bf q})}{q} \int_{\bf p} \,Q_\nu \tilde{\Lambda}^{\nu}({\bf p})\,,
\end{align}
where
\begin{equation}\label{tillambda}
    \tilde{\Lambda}^{\nu}({\bf p})=\left[\Lambda^{\nu\alpha}_{~~\alpha}(P)-\Lambda_\alpha^{~\nu\alpha}(P)
    +\Lambda_\alpha^{~\alpha\nu}(P)\right]_{p_0=p ({\bf \hat{p}}\cdot {\bf
    \hat{q}})}\,.
\end{equation}
The tensor $\Lambda$ is defined through

\begin{equation}\label{lambdadef}
    \Lambda_{\alpha \beta
    \gamma}(P)=\frac{P_\alpha-\Sigma_\alpha(P)}{(P-\Sigma(P))^2}\,\text{Im}\,\Sigma_\beta(P)\frac{P_\gamma-\Sigma^*_\gamma(P)}{(P-\Sigma^*(P))^2} \, ,
\end{equation}
where the star indicates complex conjugation. To compute the soft photon 
rate (\ref{softphotonrate}) we must specify the anisotropic distribution 
functions we use and then numerically evaluate Eqs.~(\ref{q-self}) and 
(\ref{iPi3}) with an ultraviolet cutoff $p^*$ placed on the length of three-momentum.


\section{Results}
\label{sec:results}

The total photon rate is given by adding the hard (\ref{hardphotonrate}) and 
soft (\ref{softphotonrate}) rates.  Here we specialize to the case $\xi>0$ which 
is the relevant one for times $\tau\,\gsim$ 0.2 fm/c. For the arbitrary 
isotropic distribution functions, $f_{i,\rm iso}$, which are contracted along the 
$z$-axis, we choose Bose-Einstein and Fermi-Dirac distributions in the case of 
gluons and quark/anti-quarks, respectively.  Using these distribution functions 
we reproduce the equilibrium results which have been obtained previously 
\cite{Kapusta:1991qp,Redlich:1992fr} by taking the limit $\xi \rightarrow 0$.
For $\xi \neq 0$ we find that the soft/hard logarithms of $p^*$ cancel for all 
photon angles/energies.

\begin{figure}[t]
\begin{center}
\includegraphics[width=8.15cm]{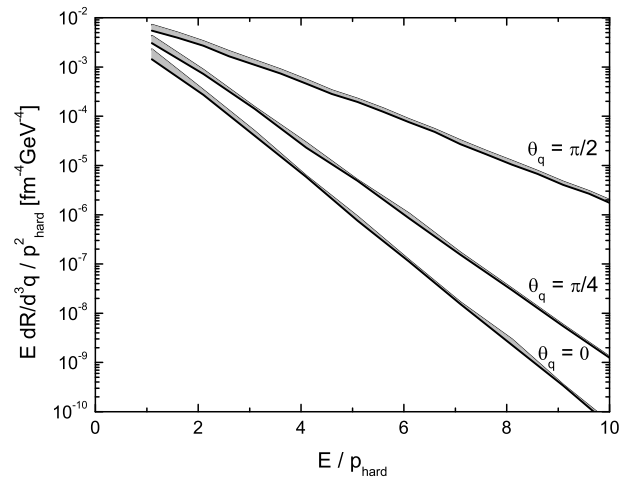}
\end{center}
\vspace{-5.5mm}
\caption{Photon rate for $\xi=10$ and $\alpha_s=0.32$ as a function of energy for 
three different photon angles, $\theta_q = \{0,\pi/4,\pi/2\}$, corresponding 
to rapidities $y = \{\infty,0.88,0\}$, respectively.}
\label{fig:qPlot}
\end{figure}

\begin{figure}[t]
\begin{center}
\includegraphics[width=8.15cm]{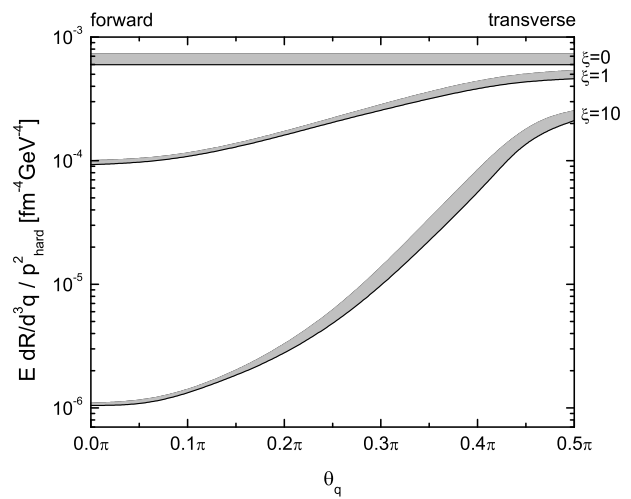}
\end{center}
\vspace{-5.5mm}
\caption{Photon rate for $\xi=\{0,1,10\}$ as a function of the photon angle,
$\theta_q$, for $E/p_{\rm hard} = 5$ and $\alpha_s=0.32$.}
\label{fig:thetaPlot}
\end{figure}

In Fig.~\ref{fig:qPlot} we plot the dependence of the total anisotropic photon 
production rate as a function of photon energy, $E$, for three different photon 
propagation angles assuming $\xi=10$ and $\alpha_s=0.32$.  This Figure shows that 
there is a clear dependence of the spectrum on photon angle with the difference 
increasing as the energy of the photon increases.  The shaded bands indicate our 
estimated theoretical uncertainty which is determined by varying the hard-soft 
separation scale $p^*$ by a factor of two around its central value which is 
parametrically $p^* \sim \sqrt{g}\,p_{\rm hard}$.  We have scaled everywhere by 
the arbitrary hard-momentum scale $p_{\rm hard}$ which appears in the quark and 
gluon distribution functions.  This scale will be time dependent with its value 
set to the nuclear saturation scale, $Q_s$, at early times and to the plasma 
temperature at late times.  For RHIC $Q_s \sim 1.4\!-\!2\;{\rm GeV}$ and 
expected initial plasma temperatures are $T_o \sim 300\!-\!400\;{\rm MeV}$.  

To compare the angular dependence at different $\xi$ in Fig.~\ref{fig:thetaPlot} 
we show the dependence of the photon production rate on angle at a fixed photon 
energy $E/p_{\rm hard}=5$ and $\alpha_s=0.32$.  As can be seen from 
Fig.~\ref{fig:thetaPlot} the difference between the forward and transverse 
production rates increases as $\xi$ increases.  To summarize the effect we 
define the photon anisotropy parameter, $ {\cal R}_\gamma \equiv 
\left(dR/d^3q|_{\theta_q=0}\right)/\left(dR/d^3q|_{\theta_q=\pi/2}\right) $. 
This ratio is one at all energies if the plasma is isotropic.  For anisotropic 
plasmas it increases as the anisotropy of the system and/or the energy of the 
photon increases.  Due to the limited experimental rapidity acceptance one could 
define this ratio at a lower angle, e.g.~$\pi/4$.  However, the ability to 
resolve anisotropies increases as the sensitivity to forward angles increases so 
it is best to compare the most forward photons possible with 
transverse photon emission.

\begin{table}[t]
\begin{tabular}{|c|c|c|c|}
\hline
${\cal R}_\gamma$ & $\;\xi=0\;$ & $\;\xi=1\;$ & $\;\xi=10\;$ \\
\hline
$\;E/p_{\rm hard}=5\;$ & 1 & 4.7 & $2\cdot10^2$ \\
$\;E/p_{\rm hard}=10\;$ & 1 & 34 & $3\cdot10^4 $ \\
\hline
\end{tabular}
\caption{Photon anisotropy parameter ${\cal R}_\gamma$ for different photon energies 
and $\xi$, assuming $\alpha_s=0.32$.}
\label{tab:kappa}
\end{table}


\section{Conclusions and Discussion}
\label{sec:conclusions}

The results presented in the previous section demonstrate that the differential 
high-energy medium photon production rate is sensitive to quark-gluon plasma 
momentum-space anisotropies with the sensitivity increasing with photon energy. 
Turning this into the final expected experimental yields as a function of photon 
energy and angle will require folding this rate together with a phenomenological 
model of the time dependence of $\xi$ and $p_{\rm hard}$. Current models of 
plasma evolution do not take into account the evolution of the momentum-space 
anisotropy of the parton distribution functions.  They therefore implicitly 
assume an isotropic thermal plasma at all times so it will be necessary to 
extend these models to include the time dependence of $\xi$ and $p_{\rm hard}$. 
However, since high-energy medium photon production is dominated by early times 
these photons will be mostly sensitive to the first 1-2 fm/c of plasma evolution 
where simple models will hopefully be efficacious.

One theoretical shortcoming of our calculation is that it does not include the 
bremsstrahlung contribution which in an equilibrium plasma also contributes at 
leading order in the coupling constant due to enhancements from collinear photon 
radiation \cite{Aurenche:1998nw,Aurenche:1999tq,Arnold:2001ba,Arnold:2001ms}. 
This contribution has been omitted because it is currently not known how to 
resolve the problem that the presence of unstable modes causes unregulated 
singularities to appear in matrix elements which involve soft gluon exchange. 
Recent work \cite{Romatschke:2006bb} has suggested that these singularities 
could be shielded by next-to-leading order corrections to the gluonic 
polarization tensor, however, the detailed evaluation of these NLO corrections 
has not been performed to date.  Absent an explicit calculation our naive 
expectation would be that these NLO corrections would yield similar angular
dependence since they are also peaked at nearly collinear angles.
\section*{Acknowledgments}
B.S. was supported by DFG Grant GR~1536/6-1.  We would also like to thank the 
Institute for Nuclear Theory, Seattle for partial support during work on this 
project.

\bibliography{anisophotons}

\begin{thebibliography}{32}
\expandafter\ifx\csname natexlab\endcsname\relax\def\natexlab#1{#1}\fi
\expandafter\ifx\csname bibnamefont\endcsname\relax
  \def\bibnamefont#1{#1}\fi
\expandafter\ifx\csname bibfnamefont\endcsname\relax
  \def\bibfnamefont#1{#1}\fi
\expandafter\ifx\csname citenamefont\endcsname\relax
  \def\citenamefont#1{#1}\fi
\expandafter\ifx\csname url\endcsname\relax
  \def\url#1{\texttt{#1}}\fi
\expandafter\ifx\csname urlprefix\endcsname\relax\def\urlprefix{URL }\fi
\providecommand{\bibinfo}[2]{#2}
\providecommand{\eprint}[2][]{\url{#2}}

\bibitem[{\citenamefont{Baier et~al.}(2001)\citenamefont{Baier, Mueller,
  Schiff, and Son}}]{Baier:2000sb}
\bibinfo{author}{\bibfnamefont{R.}~\bibnamefont{Baier}},
  \bibinfo{author}{\bibfnamefont{A.~H.} \bibnamefont{Mueller}},
  \bibinfo{author}{\bibfnamefont{D.}~\bibnamefont{Schiff}}, \bibnamefont{and}
  \bibinfo{author}{\bibfnamefont{D.~T.} \bibnamefont{Son}},
  \bibinfo{journal}{Phys. Lett.} \textbf{\bibinfo{volume}{B502}},
  \bibinfo{pages}{51} (\bibinfo{year}{2001}), \eprint{hep-ph/0009237}.

\bibitem[{\citenamefont{Mrowczynski and Thoma}(2000)}]{Mrowczynski:2000ed}
\bibinfo{author}{\bibfnamefont{S.}~\bibnamefont{Mrowczynski}} \bibnamefont{and}
  \bibinfo{author}{\bibfnamefont{M.~H.} \bibnamefont{Thoma}},
  \bibinfo{journal}{Phys. Rev.} \textbf{\bibinfo{volume}{D62}},
  \bibinfo{pages}{036011} (\bibinfo{year}{2000}), \eprint{hep-ph/0001164}.

\bibitem[{\citenamefont{Randrup and Mrowczynski}(2003)}]{Randrup:2003cw}
\bibinfo{author}{\bibfnamefont{J.}~\bibnamefont{Randrup}} \bibnamefont{and}
  \bibinfo{author}{\bibfnamefont{S.}~\bibnamefont{Mrowczynski}},
  \bibinfo{journal}{Phys. Rev.} \textbf{\bibinfo{volume}{C68}},
  \bibinfo{pages}{034909} (\bibinfo{year}{2003}), \eprint{nucl-th/0303021}.

\bibitem[{\citenamefont{Romatschke and Strickland}(2003)}]{Romatschke:2003ms}
\bibinfo{author}{\bibfnamefont{P.}~\bibnamefont{Romatschke}} \bibnamefont{and}
  \bibinfo{author}{\bibfnamefont{M.}~\bibnamefont{Strickland}},
  \bibinfo{journal}{Phys. Rev.} \textbf{\bibinfo{volume}{D68}},
  \bibinfo{pages}{036004} (\bibinfo{year}{2003}), \eprint{hep-ph/0304092}.

\bibitem[{\citenamefont{Arnold et~al.}(2003)\citenamefont{Arnold, Lenaghan, and
  Moore}}]{Arnold:2003rq}
\bibinfo{author}{\bibfnamefont{P.}~\bibnamefont{Arnold}},
  \bibinfo{author}{\bibfnamefont{J.}~\bibnamefont{Lenaghan}}, \bibnamefont{and}
  \bibinfo{author}{\bibfnamefont{G.~D.} \bibnamefont{Moore}},
  \bibinfo{journal}{JHEP} \textbf{\bibinfo{volume}{08}}, \bibinfo{pages}{002}
  (\bibinfo{year}{2003}), \eprint{hep-ph/0307325}.

\bibitem[{\citenamefont{Romatschke and Strickland}(2004)}]{Romatschke:2004jh}
\bibinfo{author}{\bibfnamefont{P.}~\bibnamefont{Romatschke}} \bibnamefont{and}
  \bibinfo{author}{\bibfnamefont{M.}~\bibnamefont{Strickland}},
  \bibinfo{journal}{Phys. Rev.} \textbf{\bibinfo{volume}{D70}},
  \bibinfo{pages}{116006} (\bibinfo{year}{2004}), \eprint{hep-ph/0406188}.

\bibitem[{\citenamefont{Mrowczynski et~al.}(2004)\citenamefont{Mrowczynski,
  Rebhan, and Strickland}}]{Mrowczynski:2004kv}
\bibinfo{author}{\bibfnamefont{S.}~\bibnamefont{Mrowczynski}},
  \bibinfo{author}{\bibfnamefont{A.}~\bibnamefont{Rebhan}}, \bibnamefont{and}
  \bibinfo{author}{\bibfnamefont{M.}~\bibnamefont{Strickland}},
  \bibinfo{journal}{Phys. Rev.} \textbf{\bibinfo{volume}{D70}},
  \bibinfo{pages}{025004} (\bibinfo{year}{2004}), \eprint{hep-ph/0403256}.

\bibitem[{\citenamefont{Arnold et~al.}(2005)\citenamefont{Arnold, Moore, and
  Yaffe}}]{Arnold:2005vb}
\bibinfo{author}{\bibfnamefont{P.}~\bibnamefont{Arnold}},
  \bibinfo{author}{\bibfnamefont{G.~D.} \bibnamefont{Moore}}, \bibnamefont{and}
  \bibinfo{author}{\bibfnamefont{L.~G.} \bibnamefont{Yaffe}},
  \bibinfo{journal}{Phys. Rev.} \textbf{\bibinfo{volume}{D72}},
  \bibinfo{pages}{054003} (\bibinfo{year}{2005}), \eprint{hep-ph/0505212}.

\bibitem[{\citenamefont{Rebhan et~al.}(2005)\citenamefont{Rebhan, Romatschke,
  and Strickland}}]{Rebhan:2005re}
\bibinfo{author}{\bibfnamefont{A.}~\bibnamefont{Rebhan}},
  \bibinfo{author}{\bibfnamefont{P.}~\bibnamefont{Romatschke}},
  \bibnamefont{and}
  \bibinfo{author}{\bibfnamefont{M.}~\bibnamefont{Strickland}},
  \bibinfo{journal}{JHEP} \textbf{\bibinfo{volume}{09}}, \bibinfo{pages}{041}
  (\bibinfo{year}{2005}), \eprint{hep-ph/0505261}.

\bibitem[{\citenamefont{Romatschke and
  Venugopalan}(2006{\natexlab{a}})}]{Romatschke:2005pm}
\bibinfo{author}{\bibfnamefont{P.}~\bibnamefont{Romatschke}} \bibnamefont{and}
  \bibinfo{author}{\bibfnamefont{R.}~\bibnamefont{Venugopalan}},
  \bibinfo{journal}{Phys. Rev. Lett.} \textbf{\bibinfo{volume}{96}},
  \bibinfo{pages}{062302} (\bibinfo{year}{2006}{\natexlab{a}}),
  \eprint{hep-ph/0510121}.

\bibitem[{\citenamefont{Schenke et~al.}(2006)\citenamefont{Schenke, Strickland,
  Greiner, and Thoma}}]{Schenke:2006xu}
\bibinfo{author}{\bibfnamefont{B.}~\bibnamefont{Schenke}},
  \bibinfo{author}{\bibfnamefont{M.}~\bibnamefont{Strickland}},
  \bibinfo{author}{\bibfnamefont{C.}~\bibnamefont{Greiner}}, \bibnamefont{and}
  \bibinfo{author}{\bibfnamefont{M.~H.} \bibnamefont{Thoma}},
  \bibinfo{journal}{Phys. Rev.} \textbf{\bibinfo{volume}{D73}},
  \bibinfo{pages}{125004} (\bibinfo{year}{2006}), \eprint{hep-ph/0603029}.

\bibitem[{\citenamefont{Manuel and Mrowczynski}(2006)}]{Manuel:2006hg}
\bibinfo{author}{\bibfnamefont{C.}~\bibnamefont{Manuel}} \bibnamefont{and}
  \bibinfo{author}{\bibfnamefont{S.}~\bibnamefont{Mrowczynski}},
  \bibinfo{journal}{Phys. Rev.} \textbf{\bibinfo{volume}{D74}},
  \bibinfo{pages}{105003} (\bibinfo{year}{2006}), \eprint{hep-ph/0606276}.

\bibitem[{\citenamefont{Romatschke and
  Venugopalan}(2006{\natexlab{b}})}]{Romatschke:2006nk}
\bibinfo{author}{\bibfnamefont{P.}~\bibnamefont{Romatschke}} \bibnamefont{and}
  \bibinfo{author}{\bibfnamefont{R.}~\bibnamefont{Venugopalan}},
  \bibinfo{journal}{Phys. Rev.} \textbf{\bibinfo{volume}{D74}},
  \bibinfo{pages}{045011} (\bibinfo{year}{2006}{\natexlab{b}}),
  \eprint{hep-ph/0605045}.

\bibitem[{\citenamefont{Romatschke and Rebhan}(2006)}]{Romatschke:2006wg}
\bibinfo{author}{\bibfnamefont{P.}~\bibnamefont{Romatschke}} \bibnamefont{and}
  \bibinfo{author}{\bibfnamefont{A.}~\bibnamefont{Rebhan}}
  (\bibinfo{year}{2006}), \eprint{hep-ph/0605064}.

\bibitem[{\citenamefont{Dumitru et~al.}(2006)\citenamefont{Dumitru, Nara, and
  Strickland}}]{Dumitru:2006pz}
\bibinfo{author}{\bibfnamefont{A.}~\bibnamefont{Dumitru}},
  \bibinfo{author}{\bibfnamefont{Y.}~\bibnamefont{Nara}}, \bibnamefont{and}
  \bibinfo{author}{\bibfnamefont{M.}~\bibnamefont{Strickland}}
  (\bibinfo{year}{2006}), \eprint{hep-ph/0604149}.

\bibitem[{\citenamefont{Shuryak}(1978)}]{Shuryak:1978ij}
\bibinfo{author}{\bibfnamefont{E.~V.} \bibnamefont{Shuryak}},
  \bibinfo{journal}{Phys. Lett.} \textbf{\bibinfo{volume}{B78}},
  \bibinfo{pages}{150} (\bibinfo{year}{1978}).

\bibitem[{\citenamefont{Kajantie and Miettinen}(1981)}]{Kajantie:1981wg}
\bibinfo{author}{\bibfnamefont{K.}~\bibnamefont{Kajantie}} \bibnamefont{and}
  \bibinfo{author}{\bibfnamefont{H.~I.} \bibnamefont{Miettinen}},
  \bibinfo{journal}{Zeit. Phys.} \textbf{\bibinfo{volume}{C9}},
  \bibinfo{pages}{341} (\bibinfo{year}{1981}).

\bibitem[{\citenamefont{Kapusta et~al.}(1991)\citenamefont{Kapusta, Lichard,
  and Seibert}}]{Kapusta:1991qp}
\bibinfo{author}{\bibfnamefont{J.~I.} \bibnamefont{Kapusta}},
  \bibinfo{author}{\bibfnamefont{P.}~\bibnamefont{Lichard}}, \bibnamefont{and}
  \bibinfo{author}{\bibfnamefont{D.}~\bibnamefont{Seibert}},
  \bibinfo{journal}{Phys. Rev.} \textbf{\bibinfo{volume}{D44}},
  \bibinfo{pages}{2774} (\bibinfo{year}{1991}).

\bibitem[{\citenamefont{Redlich et~al.}(1992)\citenamefont{Redlich, Baier,
  Nakkagawa, and Niegawa}}]{Redlich:1992fr}
\bibinfo{author}{\bibfnamefont{K.}~\bibnamefont{Redlich}},
  \bibinfo{author}{\bibfnamefont{R.}~\bibnamefont{Baier}},
  \bibinfo{author}{\bibfnamefont{H.}~\bibnamefont{Nakkagawa}},
  \bibnamefont{and} \bibinfo{author}{\bibfnamefont{A.}~\bibnamefont{Niegawa}},
  \bibinfo{journal}{Nucl. Phys.} \textbf{\bibinfo{volume}{A544}},
  \bibinfo{pages}{511} (\bibinfo{year}{1992}).

\bibitem[{\citenamefont{Aurenche et~al.}(1998)\citenamefont{Aurenche, Gelis,
  Kobes, and Zaraket}}]{Aurenche:1998nw}
\bibinfo{author}{\bibfnamefont{P.}~\bibnamefont{Aurenche}},
  \bibinfo{author}{\bibfnamefont{F.}~\bibnamefont{Gelis}},
  \bibinfo{author}{\bibfnamefont{R.}~\bibnamefont{Kobes}}, \bibnamefont{and}
  \bibinfo{author}{\bibfnamefont{H.}~\bibnamefont{Zaraket}},
  \bibinfo{journal}{Phys. Rev.} \textbf{\bibinfo{volume}{D58}},
  \bibinfo{pages}{085003} (\bibinfo{year}{1998}), \eprint{hep-ph/9804224}.

\bibitem[{\citenamefont{Aurenche et~al.}(2000)\citenamefont{Aurenche, Gelis,
  and Zaraket}}]{Aurenche:1999tq}
\bibinfo{author}{\bibfnamefont{P.}~\bibnamefont{Aurenche}},
  \bibinfo{author}{\bibfnamefont{F.}~\bibnamefont{Gelis}}, \bibnamefont{and}
  \bibinfo{author}{\bibfnamefont{H.}~\bibnamefont{Zaraket}},
  \bibinfo{journal}{Phys. Rev.} \textbf{\bibinfo{volume}{D61}},
  \bibinfo{pages}{116001} (\bibinfo{year}{2000}), \eprint{hep-ph/9911367}.

\bibitem[{\citenamefont{Arnold et~al.}(2001{\natexlab{a}})\citenamefont{Arnold,
  Moore, and Yaffe}}]{Arnold:2001ba}
\bibinfo{author}{\bibfnamefont{P.}~\bibnamefont{Arnold}},
  \bibinfo{author}{\bibfnamefont{G.~D.} \bibnamefont{Moore}}, \bibnamefont{and}
  \bibinfo{author}{\bibfnamefont{L.~G.} \bibnamefont{Yaffe}},
  \bibinfo{journal}{JHEP} \textbf{\bibinfo{volume}{11}}, \bibinfo{pages}{057}
  (\bibinfo{year}{2001}{\natexlab{a}}), \eprint{hep-ph/0109064}.

\bibitem[{\citenamefont{Arnold et~al.}(2001{\natexlab{b}})\citenamefont{Arnold,
  Moore, and Yaffe}}]{Arnold:2001ms}
\bibinfo{author}{\bibfnamefont{P.}~\bibnamefont{Arnold}},
  \bibinfo{author}{\bibfnamefont{G.~D.} \bibnamefont{Moore}}, \bibnamefont{and}
  \bibinfo{author}{\bibfnamefont{L.~G.} \bibnamefont{Yaffe}},
  \bibinfo{journal}{JHEP} \textbf{\bibinfo{volume}{12}}, \bibinfo{pages}{009}
  (\bibinfo{year}{2001}{\natexlab{b}}), \eprint{hep-ph/0111107}.

\bibitem[{\citenamefont{Braaten and Yuan}(1991)}]{Braaten:1991dd}
\bibinfo{author}{\bibfnamefont{E.}~\bibnamefont{Braaten}} \bibnamefont{and}
  \bibinfo{author}{\bibfnamefont{T.~C.} \bibnamefont{Yuan}},
  \bibinfo{journal}{Phys. Rev. Lett.} \textbf{\bibinfo{volume}{66}},
  \bibinfo{pages}{2183} (\bibinfo{year}{1991}).

\bibitem[{\citenamefont{Keldysh}(1964)}]{Ke64}
\bibinfo{author}{\bibfnamefont{L.}~\bibnamefont{Keldysh}},
  \bibinfo{journal}{Zh. Eks. Teor. Fiz.} \textbf{\bibinfo{volume}{47}},
  \bibinfo{pages}{1515} (\bibinfo{year}{1964}).

\bibitem[{\citenamefont{Keldysh}(1965)}]{Ke65}
\bibinfo{author}{\bibfnamefont{L.}~\bibnamefont{Keldysh}},
  \bibinfo{journal}{Sov. Phys. JETP} \textbf{\bibinfo{volume}{20}},
  \bibinfo{pages}{1018} (\bibinfo{year}{1965}).

\bibitem[{\citenamefont{Chou et~al.}(1985)\citenamefont{Chou, Su, Hao, and
  Yu}}]{Chou:1984es}
\bibinfo{author}{\bibfnamefont{K.-c.} \bibnamefont{Chou}},
  \bibinfo{author}{\bibfnamefont{Z.-b.} \bibnamefont{Su}},
  \bibinfo{author}{\bibfnamefont{B.-l.} \bibnamefont{Hao}}, \bibnamefont{and}
  \bibinfo{author}{\bibfnamefont{L.}~\bibnamefont{Yu}}, \bibinfo{journal}{Phys.
  Rept.} \textbf{\bibinfo{volume}{118}}, \bibinfo{pages}{1}
  (\bibinfo{year}{1985}).

\bibitem[{\citenamefont{Mrowczynski and Heinz}(1994)}]{Mrowczynski:1992hq}
\bibinfo{author}{\bibfnamefont{S.}~\bibnamefont{Mrowczynski}} \bibnamefont{and}
  \bibinfo{author}{\bibfnamefont{U.~W.} \bibnamefont{Heinz}},
  \bibinfo{journal}{Ann. Phys.} \textbf{\bibinfo{volume}{229}},
  \bibinfo{pages}{1} (\bibinfo{year}{1994}).

\bibitem[{\citenamefont{Calzetta and Hu}(1988)}]{Calzetta:1986cq}
\bibinfo{author}{\bibfnamefont{E.}~\bibnamefont{Calzetta}} \bibnamefont{and}
  \bibinfo{author}{\bibfnamefont{B.~L.} \bibnamefont{Hu}},
  \bibinfo{journal}{Phys. Rev.} \textbf{\bibinfo{volume}{D37}},
  \bibinfo{pages}{2878} (\bibinfo{year}{1988}).

\bibitem[{\citenamefont{Baier et~al.}(1997)\citenamefont{Baier, Dirks, Redlich,
  and Schiff}}]{Baier:1997xc}
\bibinfo{author}{\bibfnamefont{R.}~\bibnamefont{Baier}},
  \bibinfo{author}{\bibfnamefont{M.}~\bibnamefont{Dirks}},
  \bibinfo{author}{\bibfnamefont{K.}~\bibnamefont{Redlich}}, \bibnamefont{and}
  \bibinfo{author}{\bibfnamefont{D.}~\bibnamefont{Schiff}},
  \bibinfo{journal}{Phys. Rev.} \textbf{\bibinfo{volume}{D56}},
  \bibinfo{pages}{2548} (\bibinfo{year}{1997}), \eprint{hep-ph/9704262}.

\bibitem[{\citenamefont{Schenke and Strickland}(2006)}]{Schenke:2006fz}
\bibinfo{author}{\bibfnamefont{B.}~\bibnamefont{Schenke}} \bibnamefont{and}
  \bibinfo{author}{\bibfnamefont{M.}~\bibnamefont{Strickland}},
  \bibinfo{journal}{Phys. Rev.} \textbf{\bibinfo{volume}{D74}},
  \bibinfo{pages}{065004} (\bibinfo{year}{2006}), \eprint{hep-ph/0606160}.

\bibitem[{\citenamefont{Romatschke}(2006)}]{Romatschke:2006bb}
\bibinfo{author}{\bibfnamefont{P.}~\bibnamefont{Romatschke}}
  (\bibinfo{year}{2006}), \eprint{hep-ph/0607327}.

\end{thebibliography}

\end{document}